\documentclass[letter paper, 10 pt, conference]{ieeeconf}

\IEEEoverridecommandlockouts                              

\usepackage[utf8]{inputenc}
\usepackage{amsmath, amsfonts, amssymb}
\usepackage{tikz}
\usepackage{cleveref}
\usepackage[font=small]{caption}
\usetikzlibrary{arrows,shapes,automata,backgrounds,petri,calc}
\usepackage{cleveref}
\usepackage{subfig}
\usepackage{epstopdf}
\usepackage{epsfig}
\usepackage{wrapfig}
\usepackage[textsize= scriptsize]{todonotes}

\newcommand{\R}{\mathbb{R}}

\newcommand{\E}{\mathbb{E}}

\newcommand{\K}{\mathcal{K}}
\newcommand{\Q}{\mathcal{Q}}
\newcommand{\Qw}{\mathcal{Q}^w}

\newcommand{\Kd}{\K_{\rm DMD}}
\newcommand{\Kdr}{\K_{{\rm DMD}, r}}
\newcommand{\Kdt}{\tilde{\K}_{\rm DMD}}
\newcommand{\Kdrt}{\tilde{\K}_{{\rm DMD}, r}}

\newcommand{\Kdtr}{\tilde{\K}_{\rm DMD}^{\rm reg}}

\renewcommand{\baselinestretch}{0.997}

\DeclareMathOperator{\argmin}{\arg\!\min}

\def\mf{\mathbf}
\def\mb{\mathbb}
\def\mc{\mathcal}
\def\beq{\begin{equation*}}
\def\eeq{\end{equation*}}
\def\bql{\begin{equation}}
\def\eql{\end{equation}}
\def\bqn{\begin{eqnarray*}}
\def\eqn{\end{eqnarray*}}
\def\bnl{\begin{eqnarray}}
\def\enl{\end{eqnarray}}
\def\bna{\bql\begin{array}{rcl}}
\def\ena{\end{array}\eql}
\def\bnn{\beq\begin{array}{rcl}}
\def\enn{\end{array}\eeq}
\def\bma{\begin{bmatrix}}
\def\ema{\end{bmatrix}}
\def\bmx{\begin{matrix}}
\def\emx{\end{matrix}}
\def\ben{\begin{enumerate}}
\def\een{\end{enumerate}}
\def\bit{\begin{itemize}}
\def\eit{\end{itemize}}
\def\bei{\begin{itemize}}
\def\eei{\end{itemize}}
\def\bet{\begin{tabular}}
\def\eet{\end{tabular}}

\newtheorem{thm}{Theorem}
\newtheorem{lm}{ Lemma}

\newtheorem{pr}{ Proposition}
\newtheorem{rem}{ Remark}

\newtheorem{corr}{ Corollary}

\newtheorem{asm}{Assumption}

\title{\LARGE\bf On the Effect of Quantization on Dynamic Mode Decomposition}
\author{ Dipankar Maity, Debdipta Goswami, and Sriram Narayanan 
\thanks{The research of D. Maity was supported by the grant ARL DCIST CRA W911NF-17-2-0181.}
\thanks{D. Maity is with the Department of Electrical and Computer Engineering and an affiliated faculty of the North Carolina Battery Complexity, Autonomous Vehicle, and Electrification Research Center (BATT CAVE), University of North Carolina at Charlotte,  NC, 28223, USA.
Email: {\tt {dmaity@charlotte.edu}}
}
\thanks{
D. Goswami is with the Department of Mechanical and Aerospace Engineering, The Ohio State University, Columbus,
OH, 43210, USA. Email:{ \tt goswami.78@osu.edu }
}
\thanks{
S. Narayanan is with the Department of Mechanical and Aerospace Engineering, The Ohio State University, Columbus,
OH, 43210, USA. Email:{ \tt narayanan.121@osu.edu }
}
}

\begin{document}

\maketitle

\begin{abstract}
    Dynamic Mode Decomposition (DMD) is a widely used data-driven algorithm for estimating the Koopman Operator. 
    This paper investigates how the estimation process is affected when the data is quantized.
    Specifically, we examine the fundamental connection between estimates of the operator obtained from unquantized data and those from quantized data. 
    Furthermore, using the law of large numbers, we demonstrate that, under a \textit{large data regime}, the quantized estimate can be considered a regularized version of the unquantized estimate.
    This key theoretical finding paves the way to accurately recover the unquantized estimate from quantized data. 
    We also explore the relationship between the two estimates in the \textit{finite data regime}.
    The theory is validated through repeated numerical experiments conducted on three different dynamical systems.
\end{abstract}

\section{Introduction}


Koopman operator theory has found widespread applications in various fields such as fluid mechanics \cite{Rowley2009}, plasma dynamics \cite{nayak2021koopman}, control systems \cite{Otto2020}, unmanned aircraft systems \cite{narayanan2023}, and traffic prediction \cite{Avila2020}. In addition, it is also being used for machine learning tasks and training deep neural networks \cite{Dogra2020}.
One area of active research is the finite-dimensional estimation of the Koopman operator using data-driven methods.
Extended Dynamic Mode Decomposition (EDMD) is a prominent data-driven estimation technique that involves solving a least-square problem using data snapshots from the dynamical system \cite{Williams2015}.
It is well-understood that the quality of the Koopman operator estimate improves/degrades with an increase/decrease in the amount of data, as expected \cite{Arbabi2017, Hirsh2020, Lee2024}.
On the other hand, it is not clear how the quality of the data affects the estimation process.

Researchers has primarily focused on developing new data-driven methods or improving the existing ones for estimating the Koopman operator. 
It has been implicitly assumed that the systems implementing these algorithms have ample resources for handling large datasets, which becomes a significant concern when these data-intensive algorithms are implemented on resource limited systems, such as low-powered light-weight robotics applications \cite{Folkestad2022, Cleary2020}.  
While implementing these methods on to systems with limited physical resources (e.g., memory, bandwidth etc.), we may need to suitably modify them in a resource-aware fashion to make them compatible with the underlying hardware and resource constraints. 
These resource constraints may necessitate the data to be quantized, where the qunatization resolution depends on the available resources.

Quantization is a natural remedy to communication constraints in practical implementations. 
This is particularly applicable to cyber-physical-systems, multi-agent systems, and networked control systems where communication among systems/components are necessary to receive data from distributed sensors. 
Other systems may themselves deploy a low data-length hardware where the data is quantized due to the computational limitations \cite{hougen2000miniature, gholami2021survey, li2024evaluating}. 
Therefore, it is natural to ask how the quantization process (i.e., the quality of the data) affects the estimation quality of the Koopman operator.
Furthermore, one might also ask if there are any quantization schemes that are particularly suitable in this context.

To that end, we study the effects of \textit{dither quantization} \cite{gray1993dithered} -- a highly effective and widely used quantization scheme in communications and signal processing -- on the EDMD method. 
The contributions of this work are: (1) We address the fundamental question of whether and how one may recover the original solution (i.e., the one obtained from unquantized data) even with quantized data.
Using the law of large numbers, we prove in \Cref{thm:equivalence} that the estimation under quantized data is equivalent to a regularized estimation under the unquantized data, when we have a large number of data snapshots. 
We show the connection between the regularization parameter and the quantization resolution. 
(2) We then further investigate this connection inn the small data regime as well.
In that case also, we are able to analytically show how the difference between the estimates depends on the quantization resolution. 
(3) We validate our theory on three different systems with repeated experimentation under different quantization resolutions. 

The rest of the paper is organized as follows: We provide necessary background materials on Koopman Operator theory, Dynamic Mode Decomposition, and Dither Quantization in \Cref{sec:background}. 
We define our problem statement and research objectives in \Cref{sec:ProblemStatement}. 
We analyze the dither quantized dynamic mode decomposition (DQDMD) in \Cref{sec:QuantizedDMD} and demonstrate the connection between the solution obtained from DQDMD and the regular DMD/EDMD. 
We discuss our observations from implementing DQDMD on three dynamical systems in \Cref{sec:Validation}. 
Finally, we conclude the paper in \Cref{sec:conclusions}.

\textit{Notations:} The set of non-negative integers are denoted by $\mb{N}_0$. For a matrix $M$, we denote its Moore--Penrose inverse by $M^{\dagger}$, its transpose by $M^\top$, and hermitian transpose by $M^*$. 
The Big-O notation is denoted by $O(\cdot)$.

\section{Background} \label{sec:background}
\subsection{Koopman Operator Theory}
Consider a discrete-time dynamical system on a $n$-dimensional compact manifold $\mathcal{M}$, evolving according to the flow-map ${f}:\mc{M}\mapsto \mc{M}$ as follows: 
\begin{equation} \label{Eq: Dynamics}
    {x}_{t+1} = {f}({x}_{t}),\quad {x}_t\in\mc{M},\quad t\in\mb{N}_0.
\end{equation}
Let $\mc{F}$ be a Banach space of complex-valued observables $\varphi:\mc{M}\rightarrow \mb{C}$. The discrete-time \emph{Koopman operator} $\mc{K}:\mc{F}\rightarrow \mc{F}$ is defined as
\begin{equation}
    \mc{K}\circ\varphi(\cdot) = \varphi \circ {f}(\cdot),\quad \text{with}~~\varphi({x}_{t+1})=\mc{K}\varphi({x}_{t}),
\end{equation}
where $\mc{K}$ is infinite-dimensional, and linear over its argument. The scalar observables $\varphi$ are referred to as the Koopman observables.

A Koopman eigenfunction $\phi_i$ is a special  observable that satisfies $(\mc{K}\phi_i)(\cdot)=\lambda_i \phi_i(\cdot)$, for some eigenvalue $\lambda_i \in \mb{C}$. 
Considering the Koopman eigenfunctions (i.e., $\{\phi_i\}_{i \in \mb{N}}$) span the Koopman observables,  any {vector valued observable ${g}\in\mc{F}^p=[\varphi_1,~\varphi_2,~\ldots,~\varphi_p]^\top$} can be expressed as a sum of Koopman eigenfunctions ${g}(\cdot)=\sum_{i=1}^{\infty}\phi_i(\cdot){v}^{{g}}_i$, where ${v}^{{g}}_i\in\mb{R}^p, i=1,2,\ldots,$ are called the \emph{Koopman modes} of the observable $g(\cdot)$. This modal decomposition provides the growth/decay rate $|\lambda_i|$ and frequency $\angle{\lambda_i}$ of different Koopman modes via its time evolution:
\begin{equation}\label{eq:koop_decomp}
    {g}({x}_t) = \sum_{i=1}^{\infty}\lambda_i^t\phi_i({x}_0){v}^{{g}}_i.
\end{equation}
\noindent The Koopman eigenvalues ($\lambda_i$) and eigenfunctions ($\phi_i$) are properties of the dynamics only, whereas the Koopman modes ($v^i_g$) depend on the observable ($g$). 


Several methods have also been developed to compute the Koopman modal decomposition, e.g., DMD and EDMD \cite{schmid2010, Williams2015}, Ulam-Galerkin methods, and deep neural networks \cite{otto2019linearly, Yeung2019}. 
In this work, we focus on the EDMD method, which is briefly described below. 

\subsection{Approximation of Koopman Operator: Dynamic Mode Decomposition}
DMD is a data-driven method for extracting temporal feature from a sequence of time-series data using matrix factorization. Initially, DMD was introduced within the fluid dynamics community as a means of extracting spatiotemporal coherent structures from intricate flows \cite{schmid2010}. Subsequently, it was demonstrated that the spatiotemporal modes derived through DMD exhibit convergence to the Koopman modes when applied to a specific set of linear observables  \cite{Rowley2009}. This characteristic of the DMD has made it a primary computational tool for Koopman theory. DMD assumes that state variables themselves serve as the set of observables $g(x) = x \in \mb{R}^n$ \cite{Rowley2009}. 
DMD requires a pair of snapshot matrices in order to generate a linear model approximating the desired dynamical system. They are created by sampling the  state variables $x\in\mb\R^n$ at a sequence of time instants (snapshots) and concatenating them to form snapshot matrices $\mf{X} \in \mb{R}^{n\times T}$ and $\mf{X}' \in \mb{R}^{n\times T}$, where $\mf{X}'$ is one time snapshot ahead of the original snapshot matrix $\mf{X}$, i.e.,
\begin{align}
    \mf{X} = 
    \begin{bmatrix}
        {x}_0 & ... & {x}_{T-1}\\
    \end{bmatrix},
    \quad
    \mf{X}' = 
    \begin{bmatrix}\label{eq:dmd_snap_mat}
        {x}_{1} &  ... & {x}_{T}\\
    \end{bmatrix}.
\end{align}
The DMD algorithm aims to find the best linear operator $\Kd$ that relates the two snapshot matrices $\mf{X}$ and $\mf{X}'$ in a least-square sense, i.e.,
\begin{align}\label{eq:dmd_linear_fit}
    \mf{X}' &\approx {\Kd}\mf{X},
\end{align}
where $\Kd =\argmin_{A\in\R^{n\times n}}\|\mf{X}' - {A}\mf{X}\|^2$.
The algorithm efficiently extracts the eigenvalues and Koopman modes of the Koopman operator by determining the eigenvalues and eigenvectors
of $\Kd$.
The $\Kd$ matrix represents the Koopman operator in the newly mapped linear space of finite-dimensional observables.
DMD is performed by computing the singular-value decomposition (SVD) of $\mf{X}$, and then it is used for the prediction of $x_t$:
\bnl\label{Eq: DMD}
\begin{split}
 \mf{X} &= U {\Sigma} {V}^{*} \\
 \Kd &= \mf{X}' \mf{X}^{\dagger} = \mf{X}' {V} \Sigma^{-1 } {U}^{*} \\
 \hat{x}_t &= (\Kd)^t x_0,
 \end{split}
 \enl
where $U, \Sigma$ and $V$ are found from performing SVD on the data matrix $\mf{X}$.

\begin{rem}
    When the data matrix $\mf{X}$ has full row rank, $\mf{X}^\dagger$ has the closed form expression $\mf{X}^\top (\mf{X}\mf{X}^\top)^{-1}$.
    In that case, we may write $\Kd = \mf{X}' \mf{X}^\top (\mf{X}\mf{X}^\top)^{-1}$.
\end{rem}

Since the singular-values decay rapidly, first $r$ significant singular-values of $\mf{X}$ can be used for computing a reduced-order matrix $\Kdr \in \R^{n\times n}$ by projecting $\Kd$ to a lower $r$-dimnesional space as follows:
\bnl \label{Eq: ReducedDMD}
\begin{split}
 \mf{X} &= U {\Sigma} {V}^{*} \approx {U}_r {\Sigma}_r {V}^{*}_r\\
 \Kd &= \mf{X}' {V}_r \Sigma_r^{-1 } {U}_r^{*} \\
 \Kdr &= U_r^* \Kd U_r = U_r^*\mf{X}' {V}_r \Sigma_r^{-1 }
\end{split}
\enl 
We project $\Kdr$ back to the full dimension $n$ by constructing a matrix $\Upsilon$, which then can be used to predict the state:
\begin{align} \label{eq:ReducedOrderPrediction}
    \begin{split}
        \Upsilon &= \mf{X}'V_r\Sigma_r^{-1}W\\
 \hat{x}_t &= \Upsilon \Lambda^t \Upsilon^{\dagger}x_0,
    \end{split}
\end{align}
where $W$ and $\Lambda$ can be found from $\Kdr$ via its eigendecomposition $\Kdr = W\Lambda W^{-1}$.

The algorithm \eqref{Eq: DMD} or \eqref{Eq: ReducedDMD} is extended \cite{Williams2015} to a use a set of observables or dictionary functions  $\varphi(\cdot) = [\varphi^1(\cdot),\ldots,\varphi^N(\cdot)]^T: \mc{M} \mapsto \mb{C}^N $. We again define data matrices $ \Phi,  \Phi' \in \R^{N \times T}$ such that
\begin{align} \label{eq:dataMatrix}
\begin{split} 
    \Phi &= \begin{bmatrix}
         \varphi(x_0) ~  \varphi(x_1) ~ \hdots ~ \varphi(x_{T-1}) 
    \end{bmatrix},  \\
    \Phi' &= \begin{bmatrix}
         \varphi(x_1) ~  \varphi(x_2) ~ \hdots ~ \varphi(x_{T}) 
    \end{bmatrix} .
\end{split}    
\end{align} 
Now $\Phi'\approx \Kd\Phi$ is assumed and the same SVD-based methods are utilized to identify $\Kd$ or $\Kdr$, yielding the Extended Dynamic Mode Decomposition (EDMD). 
\begin{rem}
    Note that DMD is a special case of EDMD where $N=n$ and $\varphi^i(x)=x^i,\,i \in\{1,\ldots,n\}$, and $x^i$ is the $i^{\text{th}}$ component of $x$.
\end{rem}



\subsection{Dither Quantization} \label{Sec:quant}

A quantizer $q :(u_{\min},u_{\max})\subseteq \R \to \{0,\ldots, (2^b-1)\}$ is a function that maps any $x\in (u_{\min},u_{\max}) \subset \R $ to a $b$-bit binary word.
For example, a uniform quantizer takes the form 
\begin{align*}
    q(x)=\left\lfloor\frac{x-u_{\min}}{\epsilon} \right\rfloor,
\end{align*}
where 
\begin{align} \label{eq:quantizationResolution}
    \epsilon=\frac{u_{\max}-u_{\min}}{2^b}
\end{align}
denotes the quantization resolution.
Although $q(\cdot)$ is defined on the interval $(u_{\min},u_{\max})$, one may extend the definition of $q(\cdot)$ on the entire real line as follows:
\begin{align*}
    \bar{q}(x)=\begin{cases}
    q(x),\quad &x\in (u_{\min},u_{\max}),\\
    0, & x\le u_{\min},\\
    2^b-1, & x\ge  u_{\max},
    \end{cases}
\end{align*}
where $\bar q(\cdot)$ is the extended version of $q(\cdot)$. 
The region outside the interval $[u_{\min},u_{\max}]$ is the \textit{saturation} region of the quantizer $\bar{q}$.

The decoding of a mid-point uniform quantizer is performed by
\begin{align*}
    \Q(x) =\epsilon q(x) + u_{\min} + \frac{\epsilon}{2}.
\end{align*}
The quantization error is defined to be $e(x) = \mathcal{Q}(x) - x$. 
For all $x\in (u_{\min}, u_{\max})$, we have $|e(x)| \le \frac{\epsilon}{2} $.
The distribution of the quatization error plays an important role in analyzing the performance of a system employing quantization. 
The distribution of this error is correlated with the distribution of the source signal $x$.
This correlation often results in poor performance, besides making the analysis of such systems complicated. 
It has been well-established that dither quantization leads to a better performance, as demonstrated in the very first work on TV communication \cite{roberts1962picture} as well as in applications to controls \cite{stanford1960linearization}. 
Since then, a significant amount of research has been devoted in \textit{dither} quantization. 

\textit{Dither} quantization prescribes adding a noise $w$ to the source signal $x$ prior to quantization and subtract that noise during decoding \cite{gray1993dithered}, which yields the decoded signal to be $\Q(x+w) - w$. 
Thus, the quantization error becomes 
\begin{align}
    e(x) = \mathcal{Q}(x+w) - w - x.
\end{align}
Under certain assumptions on the distribution of $w$, it can be shown that this new error $e(x)$ is distributionally independent of the source $x$. 
Furthermore, this error can be shown to have a uniform distribution in $[-\frac{\epsilon}{2}, \frac{\epsilon}{2}]$. 
A typical choice of $w$ is to consider an uniformly distributed random variable with support $[-\frac{\epsilon}{2}, \frac{\epsilon}{2}]$, which satisfies all the necessary and sufficient conditions to ensure that $e $ is independent of $x$ and uniformly distributed in $[-\frac{\epsilon}{2}, \frac{\epsilon}{2}]$; see \cite{gray1993dithered}. 
Throughout this work, we will consider the \textit{dither} quantization scheme. 

When $x$ is a vector, we perform the quantization and the decoding component-wise.  
For quantizing a time-varying vector-valued process $\{x_t\}_{t\ge 0}$, we will consider a time-varying vector-valued i.i.d process $\{w_t\}_{t\ge 0}$ as the \textit{dither} signal. 
Furthermore, in the subsequent sections we will use $\Qw(x)$ as a shorthand notation for $\Q(x+w)$.

\section{Problem Statement} \label{sec:ProblemStatement}



%
Our objective in this exploratory work is to understand the effects of quantization on the estimated Koopman operator. 
To that end, we assume that we have access to quantized observables, i.e., $\tilde \varphi^i(\cdot) = \Qw \circ \varphi^i (\cdot)$ for $i = 1,\ldots, N$, where $\Qw(\cdot)$ is the dither quantization-decoding operator. 
Hence, for time instance $t$, the available data is
\begin{align} \label{eq:tildeObservables}
    \tilde \varphi(x_t) = \begin{bmatrix}
        \Qw(\varphi^1(x_t)) - w^1_t \\
        \vdots\\
        \Qw(\varphi^n(x_t)) - w^n_t
    \end{bmatrix}, 
\end{align}
where $w^i_t$ is the dither signal used during  the quantization of the $i$-th observable of time $t$, and $\Qw(\varphi^i(x_t)) = \Q(\varphi^i(x_t) + w^i_t)$ for all $i = 1, \ldots, N$ and $t = 0,\ldots, T$. 

Let $\Kdt$ denote the estimate of the Koopman operator  obtained from the quantized data.
That is, 
\begin{align} \label{eq:EDMD_quantized}
      \Kdt = \argmin_{A \in \R^{N \times N}}  \| \tilde \Phi' - A \tilde \Phi\|^2,
\end{align}
where
\begin{align*}
    \tilde \Phi &= \begin{bmatrix}
        \tilde \varphi(x_0) ~ \tilde \varphi(x_1) ~ \hdots ~ \tilde \varphi(x_{T-1}) 
    \end{bmatrix},  \\
    \tilde \Phi' &= \begin{bmatrix}
        \tilde \varphi(x_1) ~ \tilde \varphi(x_2) ~ \hdots ~ \tilde \varphi(x_{T}) 
    \end{bmatrix} ,
\end{align*}
and where $\tilde{\varphi}(\cdot)$ is defined in \eqref{eq:tildeObservables}.
On the other hand, the estimate obtained from the unquantized data is
\begin{align} \label{eq:EDMD}
    \Kd = \argmin_{A \in \R^{N \times N}}  \| \Phi' - A \Phi\|^2, 
\end{align}
where $ \Phi,  \Phi' \in \R^{N \times T}$ are the data matrices defined in \eqref{eq:dataMatrix}.

Having obtained the matrices $\Kd$ and $\Kdt$ (or, their reduced order versions), we may predict the state using \eqref{Eq: DMD} (or, using \eqref{Eq: ReducedDMD} and \eqref{eq:ReducedOrderPrediction}).
In this work, we investigate the normalized estimation errors for both full  and reduced order Koopman operators, as well as we are interested in the prediction error of the reduced order model.  
That is, we quantify how $\frac{\| \Kd - \Kdt\|}{\|\Kd\|}$, $\frac{\|\Kdr - \Kdrt\|}{\|\Kdr\|}$, and $\frac{1}{T}\sum_{t=0}^{T-1} \frac{\|\hat{x}_t - x_t \|}{\|x_t\|}$ change as we vary the word length for the quantization.

In addition to quantifying the degradation due to quantization using the three aforementioned metrics, we are also interested in developing a framework where one may obtain an improved estimate, $\Kdt^*$, that is closer to $\Kd$ than $\Kdt$ is. 
In this work we discuss such a potential method for the large data regime (i.e., when $T\to \infty$).

\section{DMD with Quantized Data} \label{sec:QuantizedDMD}

\begin{asm} \label{assm:BoundedPhi}
The observables are bounded functions.
That is, for all $i$ there exists $\ell_i < u_i$ such that  $\ell_i \le \varphi^i(x) \le u_i$ for all $x\in \R^n$.  
\end{asm}

A direct consequence of this assumption\footnote{
For practical purposes, we only need that the data matrices $\Phi$ and $\Phi'$ are bounded, since the EDMD algorithm deals only with the data and not the functions. 
Therefore, the observables do not need to be bounded functions as long as the measured data is bounded. 
} is that we may assume $u_{\min} \le \varphi^i(x) \le u_{\max}$ for all $i$. 
In case $\varphi^i(\cdot)$ is not bounded in between $u_{\min} $ and $u_{\max} $, we may consider a shifted and scaled version $\bar\varphi^i(\cdot) = \frac{\varphi(\cdot) + s}{c}$ such that $\bar\varphi^i(\cdot)$ satisfies \Cref{assm:BoundedPhi}, where the shift and scaling coefficients $s,c \in \R$ are chosen appropriately.
From this point onward we will therefore use the following assumption (Assumption 1$'$)  in place of \Cref{assm:BoundedPhi} without any additional loss of generality.\\

\noindent\textbf{Assumption 1$'$.} \textit{
The observables satisfy  $u_{\min} \le \varphi^i(x) \le $ $u_{\max}$ for all $i$ and $x\in \R^n$.}  \\



    


The main result of this section is summarized in the following theorem. 
\begin{thm}\label{thm:equivalence}
    For a large $T$,  
    \begin{align} \label{eq:equivalence}
    \begin{split}
        \Kdt &= \argmin_{A \in \R^{N \times N}} \frac{1}{T} \| \tilde \Phi' - A \tilde \Phi\|^2  \\ 
        &= \argmin_{A \in \R^{N \times N}} \frac{1}{T}  \| \Phi' - A \Phi\|^2 + \frac{\epsilon^2}{12} \|A\|^2.
        \end{split}
    \end{align}
\end{thm}

\begin{proof}
    The proof is presented in \Cref{AP:thm:equivalence}.
\end{proof}

Theorem~\ref{thm:equivalence} states that the solution $\Kdt$ can be interpreted as a solution to a regularized DMD problem, where the regularization parameter depends on the resolution of the quantizer $\epsilon$.  
A consequence of \Cref{thm:equivalence} is that the solution $\Kdt$ converges to $\Kd$ as $\epsilon$ approaches to $0$. 
Recall from \eqref{eq:quantizationResolution} that the quantization resolution $\epsilon$ is coupled with the qunatization word length. 
Thus, as the number of bits $b$ increases, we obtain $\Kdt \to \Kd$, as one would expect.

\begin{rem}
    \Cref{thm:equivalence} shows the connection between $\Kd$ and $\Kdt$ via the quantization resolution $\epsilon$. 
    It is to be noted that the relationship \eqref{eq:equivalence} holds because the quantization noises are i.i.d., which is due to the fact that dither quantization is being used. 
    A similar conclusion may not hold for other forms of quantization. 
\end{rem}

\begin{rem}
    By solving the optimization on the r.h.s. of \eqref{eq:equivalence}, one obtains 
    \begin{align} \label{eq:KDt}
        \Kdt = \Phi' \Phi^\top \left( \Phi \Phi^\top + \frac{T \epsilon^2}{12} I \right)^{-1}. 
    \end{align}
    In the unquantized case (i.e., $\epsilon = 0$), where the data matrix $\Phi$ is rank deficient, one needs to use the Moore--Penrose inverse (see \eqref{Eq: DMD}).
    However, the matrix  $\Phi \Phi^\top + \frac{T \epsilon^2}{12} I$ is always invertible, thus alleviating the need for computing any pseudo-inverse. 
\end{rem}

\Cref{thm:equivalence} not only helps in identifying the relationship between $\Kd$ and $\Kdt$, but also provides a convenient framework to potentially recover $\Kd$ from the quantized data, as discussed later in \Cref{sec:regularizedDMD}.

\begin{figure*}[t]
\centering 
\subfloat[]{\includegraphics[trim=0cm 0cm 0cm 0cm, clip=true, width=0.33\textwidth]{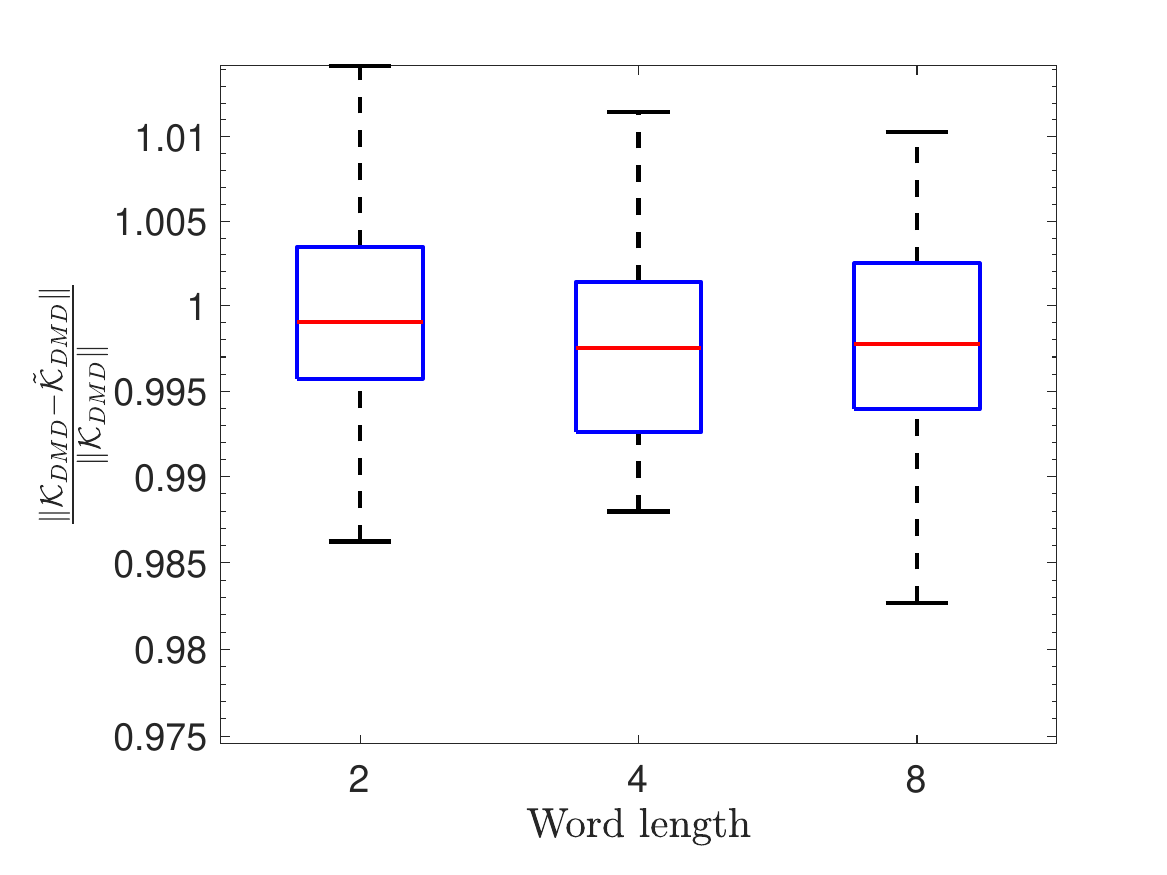}}
\subfloat[]{\includegraphics[trim=0cm 0cm 0cm 0cm, clip=true, width=0.33\textwidth]{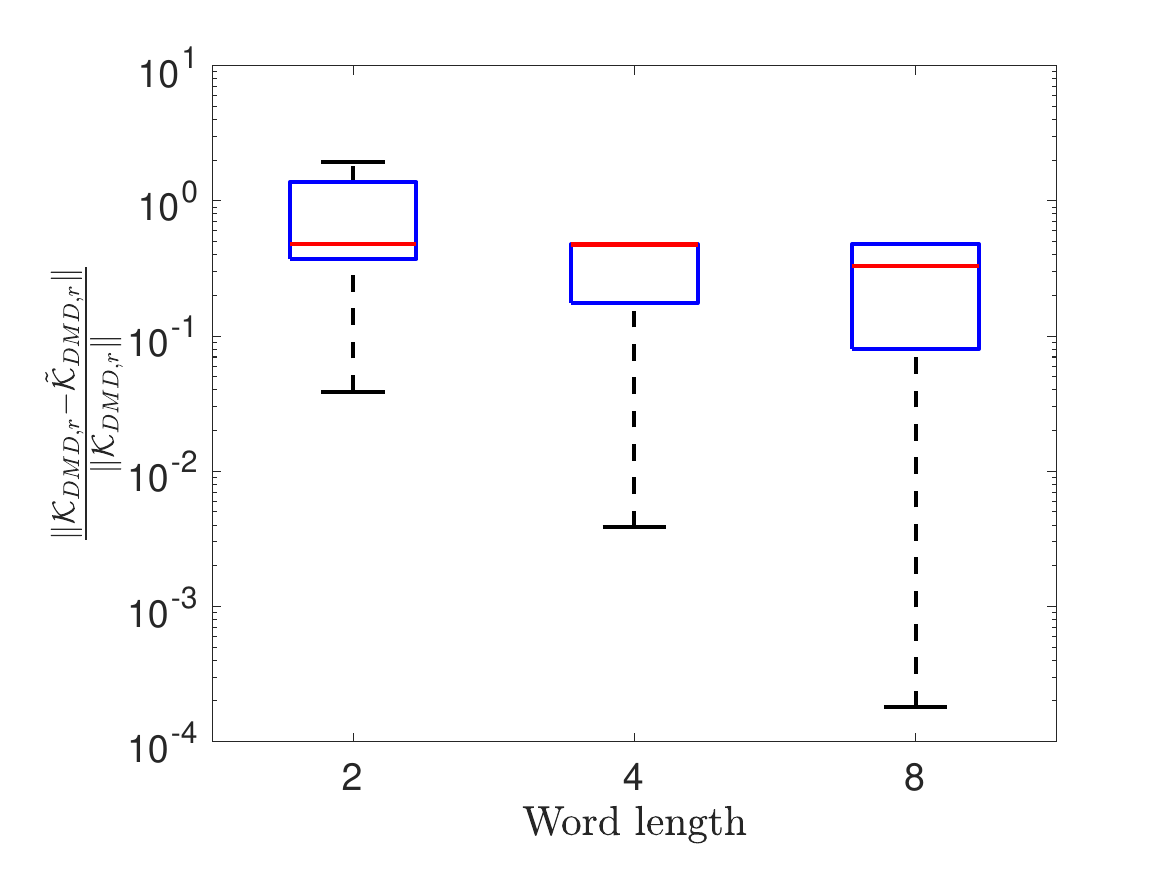}}
\subfloat[]{\includegraphics[trim=0cm 0cm 0cm 0cm, clip=true, width=0.33\textwidth]{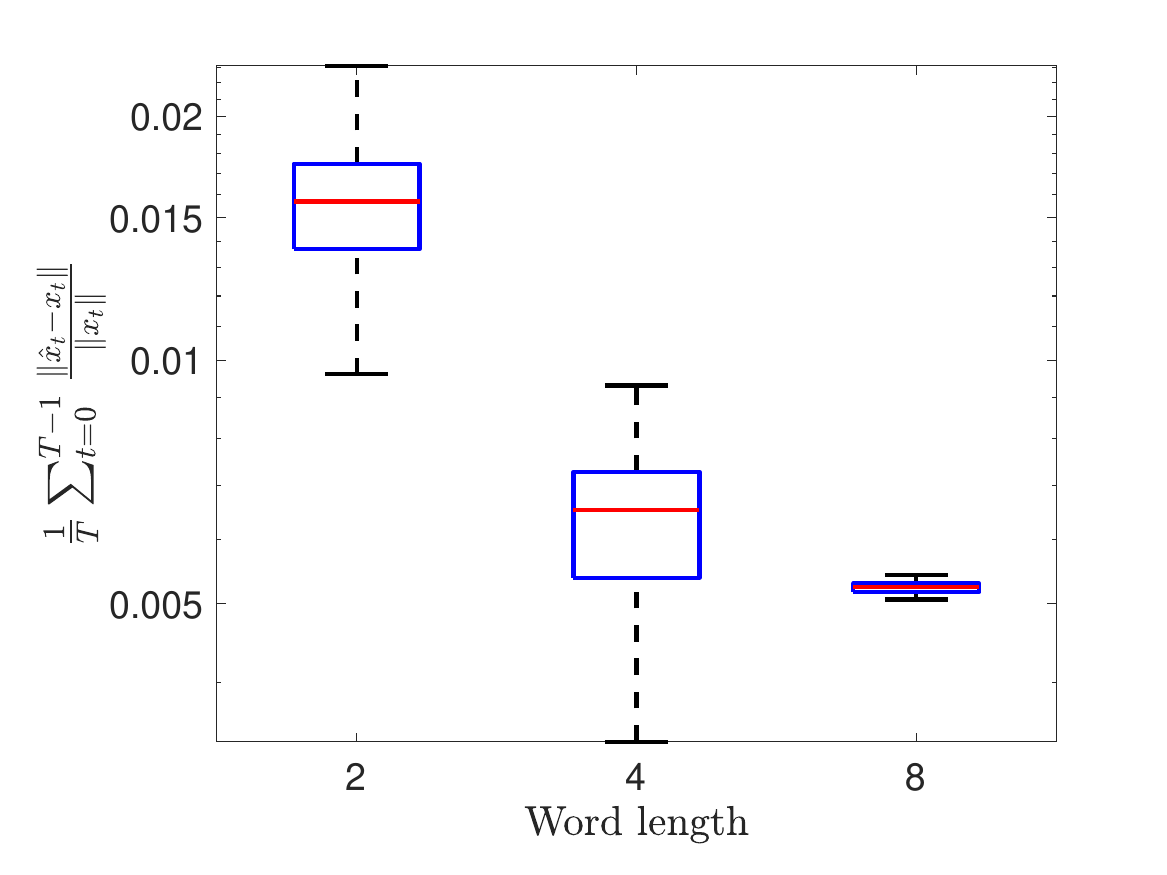}}
\caption{Error profile for negatively-damped pendulum \eqref{Eq: Pendulum}.} \label{Fig: Pendulum}
\end{figure*}
\begin{figure*}[t]
\centering 
\subfloat[]{\includegraphics[trim=0cm 0cm 0cm 0cm, clip=true, width=0.33\textwidth]{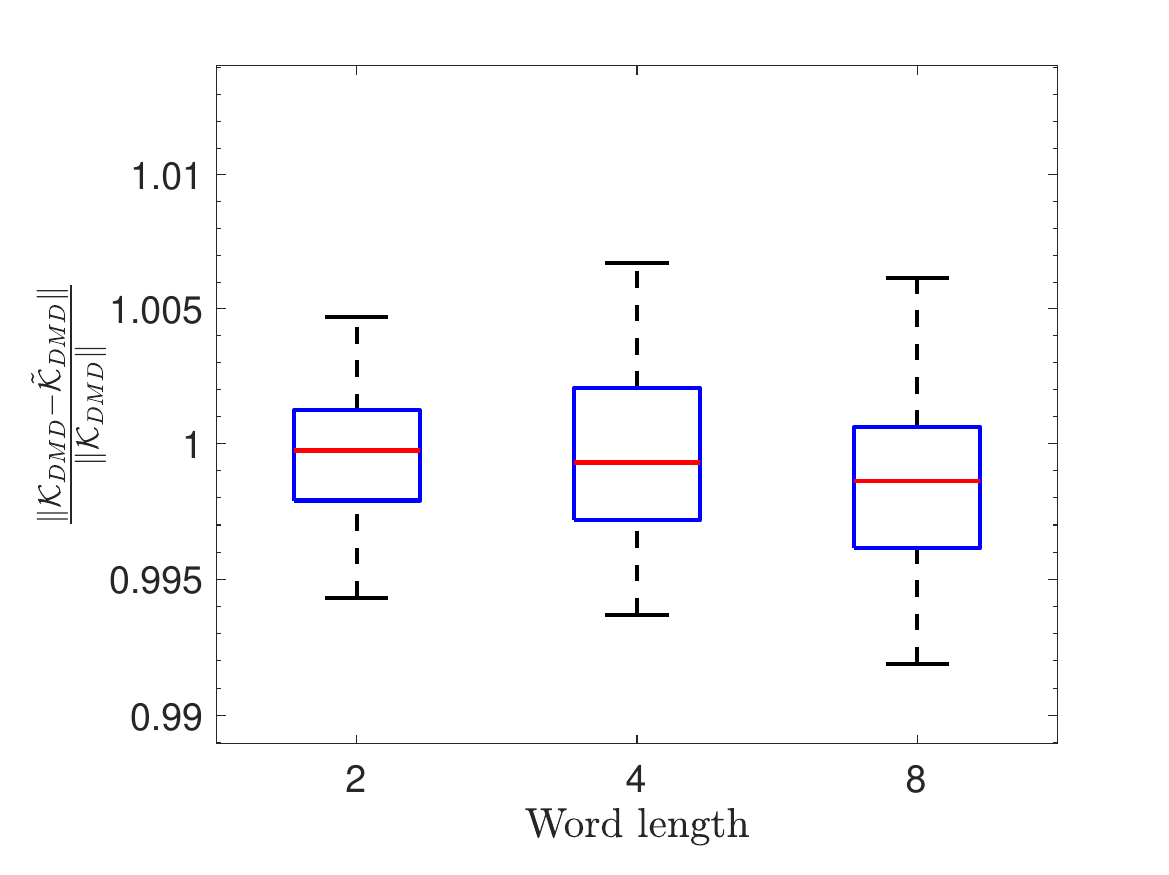}}
\subfloat[]{\includegraphics[trim=0cm 0cm 0cm 0cm, clip=true, width=0.33\textwidth]{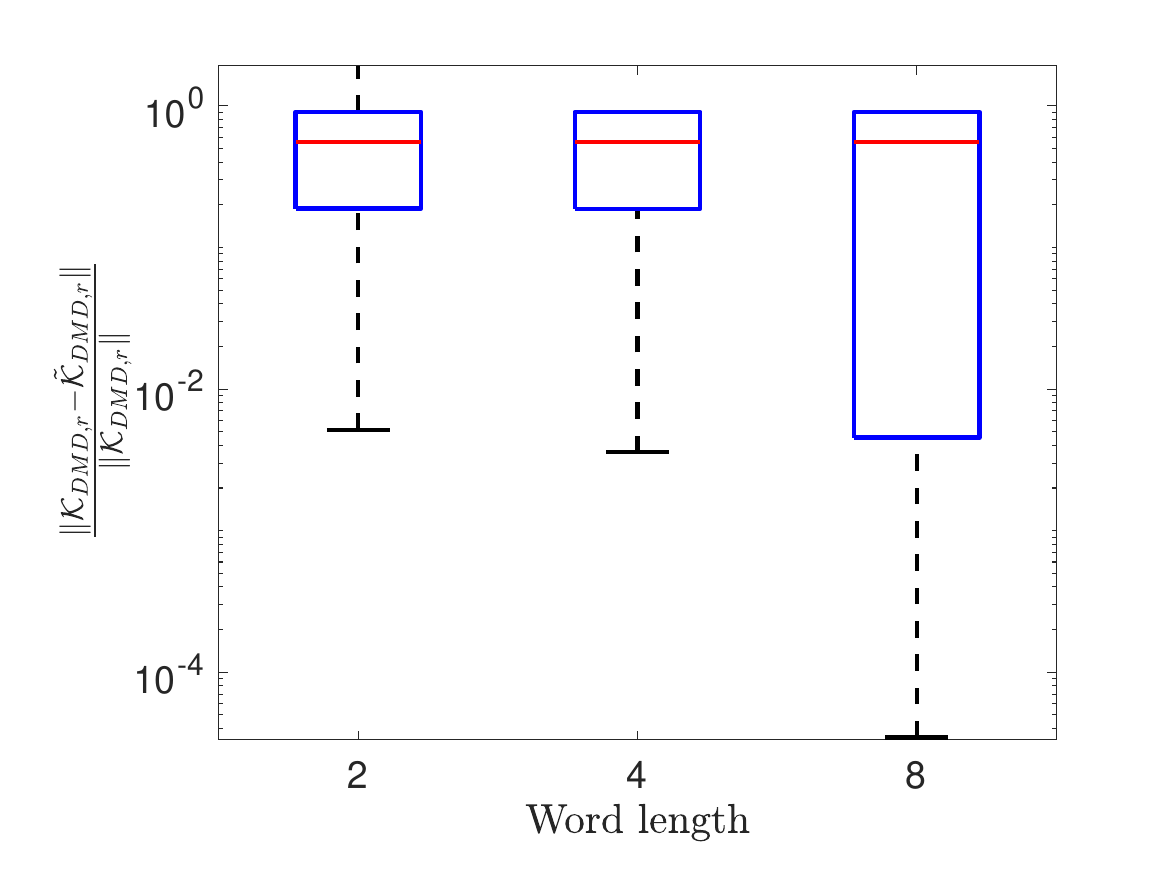}}
\subfloat[]{\includegraphics[trim=0cm 0cm 0cm 0cm, clip=true, width=0.33\textwidth]{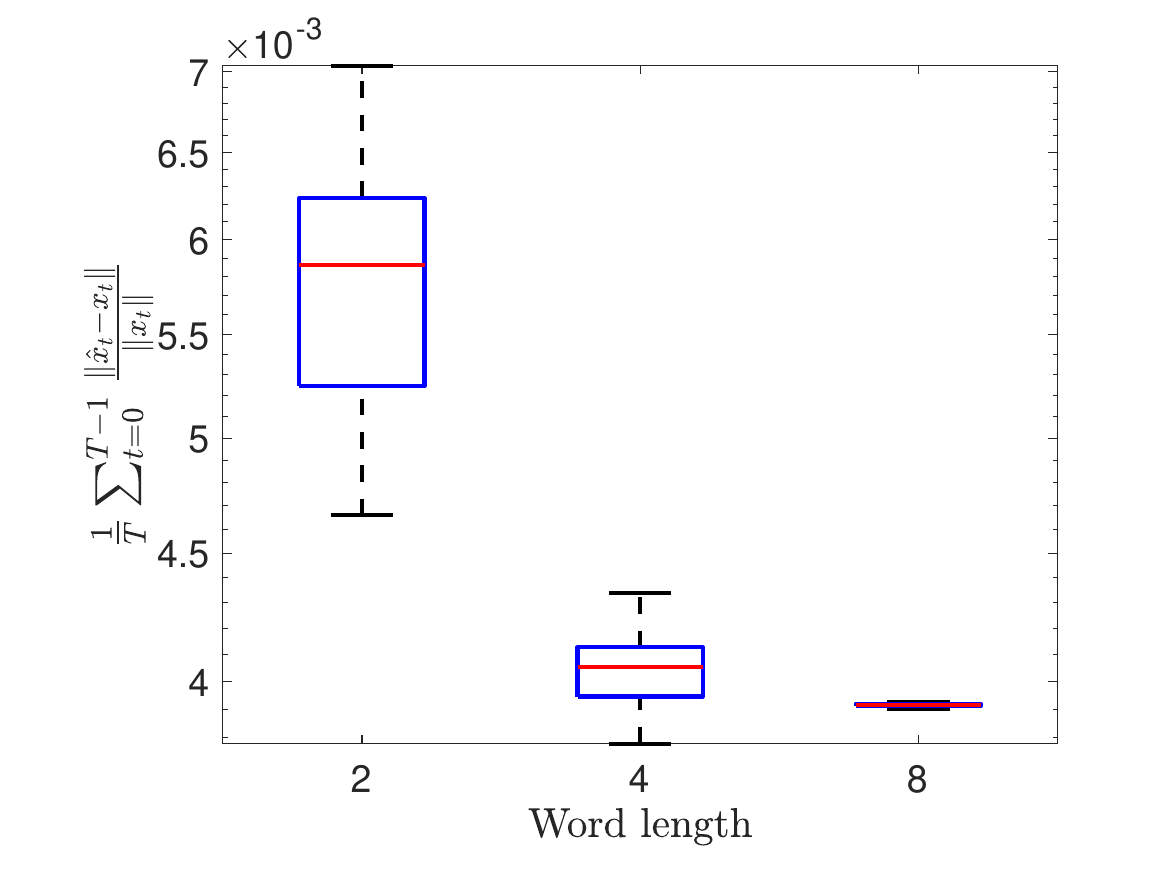}}
\caption{Error profile for Van der Pol oscillator \eqref{Eq: Vanderpol}.} \label{Fig: Van der Pol}
\end{figure*}

\subsection{Finite Data Regime} 

So far we have characterized the connection between $\Kd$ and $\Kdt$ in a large data setting, i.e., when $T$ is large (theoretically, we need $T\to \infty$). 
Next, we discuss the relationship between $\Kd$ and $\Kdt$ when $T$ is finite and potentially small (i.e., fewer time snapshots). 
The main result of this section is presented in the following theorem.

\begin{thm} \label{thm:K_epsilon}
    Let $\Phi$ and $\tilde \Phi$ be of full row rank. Then, there exists a $\K_\epsilon$ such that $\|\K_\epsilon\| = O(\epsilon)$ and 
    \begin{align}
    \Kdt = \Kd + \K_\epsilon.
    \end{align}
\end{thm}

\begin{proof}
    The proof is presented in \Cref{AP:thm:K_epsilon}.  \\
\end{proof}

The condition that $\Phi$ and $\tilde \Phi$ have full row rank in \Cref{thm:K_epsilon} is sufficient but not necessary. 
In fact, the proof may be extended without this condition through some tedious derivations. 
For the ease of the exposition, we do not delve into such details here and leave that discussion for a future work. 

Similar to \Cref{thm:equivalence}, here also we note that $\Kdt \to \Kd$ as $\epsilon \to 0$, as one would expect. 
However, \Cref{thm:K_epsilon} is a stronger result than \Cref{thm:equivalence} since it holds true for any $T$ whereas \Cref{thm:equivalence} holds when $T \to \infty$. 
The benefit of having a large $T$ in previous sections is that we may potentially recover $\Kd$ even from the quantized data, as will be discussed in \Cref{sec:regularizedDMD}.
On the other hand, we may not recover $\Kd$ from the quantized data since $\K_\epsilon$ cannot be computed without having knowledge of the unquantized data matrix $\Phi$. 

    It should be noted that $\K_\epsilon$ depends on the realization of the dither noise, and therefore, it is a random matrix. 
    We leave the investigation on the statistical properties of $\K_\epsilon$ as a potential future work.

\section{Numerical Examples} \label{sec:Validation}

The effect of dither quantization on EDMD is demonstrated on  three different systems: a simple pendulum with negative damping, Van der Pol oscillator, and fluid-flow past a cylinder. 
\subsection{Pendulum with negative damping}
A two dimensional oscillatory system with slight instability is considered as a first example.
The dynamics of a simple pendulum with a destabilizing term is described as:
\bnl \label{Eq: Pendulum}
\dot{x}_1 &=& x_2\nonumber\\
\dot{x}_2&=&0.01x_2-\sin x_1.
\enl
Since $x\in\R^2$ is very low-dimensional, we choose Hankel delay-embedding as our EDMD dictionary, i.e., we stack up $\varphi(x_t)\triangleq [x_t,\ldots,x_{t+m-1}]^\top \in\R^{2m}$ for $t=0,\ldots,T-1$. The data is generated by solving the system for $10^4$s and it is sampled at an interval $\Delta t = 0.1$s. 
We used an embedding dimension of $m=100$ and performed the DMD with dither quantization for $50$ independent Monte-Carlo trials for training length $T=500$. 
The relative 2-norm error $\frac{\| \Kd - \Kdt\|}{\|\Kd\|}$ for full-order DMD matrix, $\frac{\|\Kdr - \Kdrt\|}{\|\Kdr\|}$ for reduced order DMD matrix, and the time-average relative two norm error $\frac{1}{T}\sum_{t=0}^{T-1} \frac{\|\hat{x}_t - x_t \|}{\|x_t\|}$ between predictions using $\Kdrt$ and $\Kdr$ are shown in Fig.~\ref{Fig: Pendulum}.

We notice that the prediction error in Fig.~\ref{Fig: Pendulum}(c) decreases exponentially with the quantization word length. 
The trend is consistent in all the three subplots, where the average relative errors (shown by the red line segments) decrease with the word length. 
It is particularly interesting to see that a two-bit quantizer can provide an average normalized prediction error of less than 0.02.

\subsection{Van der Pol oscillator}
Now, we consider the limit-cyclic Van der Pol oscillator:
\bnl \label{Eq: Vanderpol}
\dot{x}_1 &=& x_2\nonumber\\
\dot{x}_2&=&(1-x_1^2)x_2-x_1,
\enl
with the same Hankel stacking $m$, sampling interval $\Delta t$, and training length $T$ to form the data matrices. The relative error metrics for this example are shown in Fig.~\ref{Fig: Van der Pol} for $50$ independent Monte-Carlo trials.
In this experiment too, we notice the same trend. 
The average normalized prediction error is of the order of $10^{-3}$ even with two bits for quantization. 
\subsection{Flow past cylinder} 
Flow past cylinder \cite{batchelor1967introduction} is another common dynamical system used for benchmarking data-driven models \cite{Otto2020, Nayak2024}. The simulation was carried out using MATLAB's FEAtool \cite{Nayak2024}. A cylinder with a diameter of $0.1$m is located at a height of $0.2$m. The fluid is characterized by its density $\rho=1$ Kg/m$^3$, and dynamic viscosity $\mu=0.001$ Kg/m s. The flow is unsteady with a maximum velocity of $1$m/s and mean velocity being $\frac{2}{3}$ of the maximum velocity. 
The simulation is run for $80$s until the steady state is achieved, and the horizontal $u$ component of the velocity is probed at every 0.02s starting from 20s. The complete dataset $\Phi\in\mb{R}^{2647\times 3001}$ consists of 3001 time samples. More details regarding the simulation setup can be found in \Cref{AP:FPC}.

\begin{figure}[t]
\centering 
\subfloat[]{\includegraphics[trim=1cm 0cm 0cm 0cm, clip=true, width=0.24\textwidth]{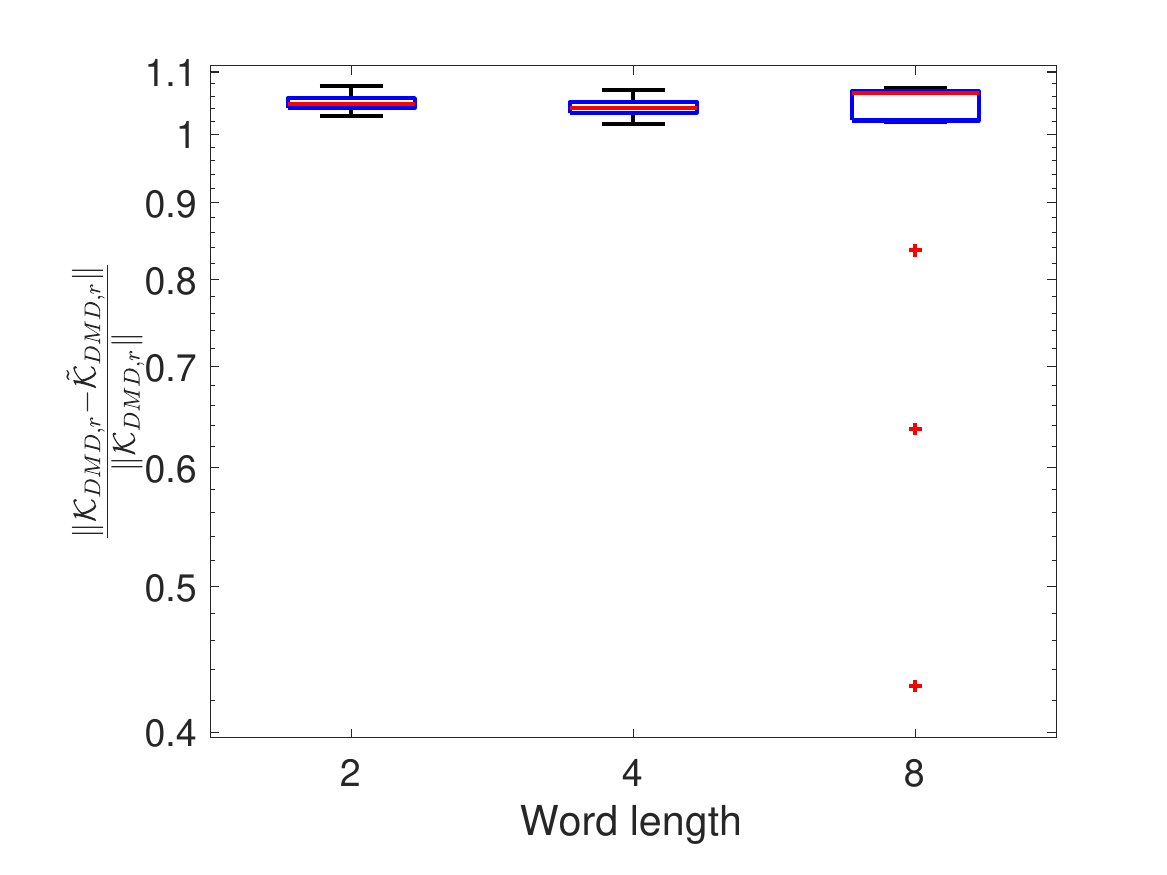}}
\subfloat[]{\includegraphics[trim=1cm 0cm 0cm 0cm, clip=true, width=0.24\textwidth]{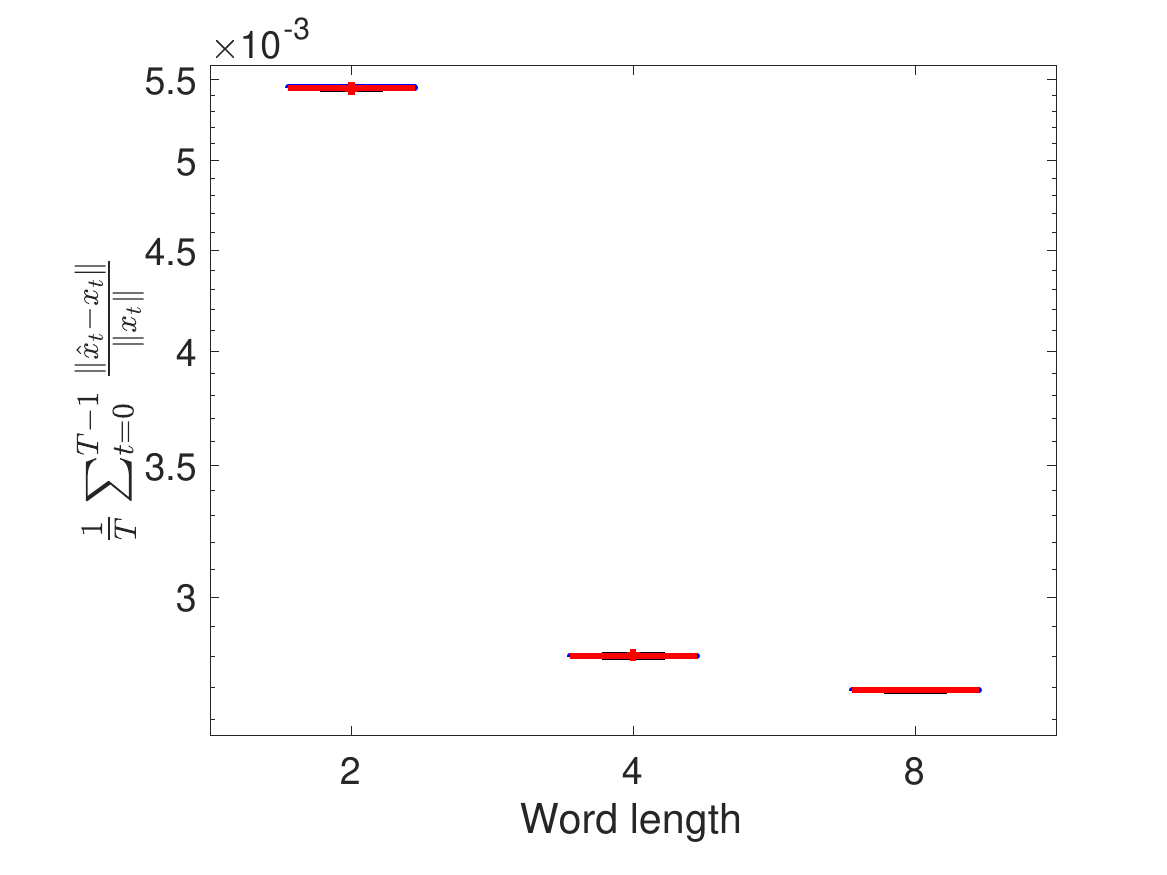}}
\caption{Error profile for flow past cylinder.} \label{Fig: FPC}
\end{figure}

Due to a large number of states, we only perform the reduced order DMD on the quantized data for 30 independent realization of dither signal and report the error metrics $\frac{\|\Kdr - \Kdrt\|}{\|\Kdr\|}$ and $\frac{1}{T}\sum_{t=0}^{T-1} \frac{\|\hat{x}_t - x_t \|}{\|x_t\|}$ in Fig.~\ref{Fig: FPC} for different word lengths

Here the normalized prediction error follows the same exponential trend with the available word length. However, the normalized matrix error metric $\frac{\|\Kdr - \Kdrt\|}{\|\Kdr\|}$ decreases at a slower rate. Also, for the 8-bit quantization, the existence of several outliers warrants larger number of Monte-Carlo runs in future.
However, these outliers are in our favor since they all produced lower relative errors.


\section{Discussions} 

\subsection{Regularized DMD} \label{sec:regularizedDMD}

Whereas a regular DMD problem prescribes \eqref{eq:EDMD_quantized}, we propose the following regularized version in presence of quantization
\begin{align} \label{eq:regularized_DMD}
   \Kdtr = \argmin_{A} \| \tilde \Phi' - A \tilde \Phi\|^2 +  \gamma \|A \|^2, 
\end{align}
where $\gamma \in \R$ is a regularization parameter.  
Our hypothesis is that, by appropriately choosing $\gamma$, one may obtain an estimate $\Kdtr$ such that $\Kdtr \approx \Kd$.

\begin{pr} \label{prop:regularizedDMD}
    For a given quantization resolution $\epsilon$, one may recover $\Kd$ from the dither quantized data, if the following regularized DMD is solved for a large $T$
    \begin{align} \label{eq:prop1}
        \frac{1}{T}\| \tilde \Phi' - A \tilde \Phi\|^2  - \frac{\epsilon^2}{12} \|A \|^2.
    \end{align}
\end{pr}

\begin{proof}
From the proof of \Cref{thm:equivalence} (see \eqref{eq:mainEquation}), we observe that, for a large enough $T$, 
\begin{align*}
     \frac{1}{T}\| \tilde \Phi' - A \tilde \Phi\|^2 = \frac{1}{T}  \|  \Phi' - A  \Phi\|^2 + \frac{\epsilon^2}{12}\|A\|^2 + \frac{N\epsilon^2}{12}.   
\end{align*}
Therefore, 
\begin{align}
     & \argmin_A   \frac{1}{T} \| \tilde \Phi' - A \tilde \Phi\|^2  - \frac{\epsilon^2}{12} \|A \|^2 \nonumber \\
    & \qquad = \argmin_A  \frac{1}{T} \|  \Phi' - A  \Phi\|^2 + \frac{N\epsilon^2}{12} = \Kd.
\end{align}
This completes the proof.
\end{proof}

\begin{rem}
    When \Cref{prop:regularizedDMD} is deployed for recovering $\Kd$ from a finite amount of data, one must be cautious about the convexity of the optimization problem in \eqref{eq:prop1}. 
    In some cases, the problem may not be convex due to the term $- \frac{\epsilon^2}{12} \|A \|^2$ and thus the optimization becomes futile. 
    In such cases, one may attempt to use a regularizer $\gamma \in ( - \frac{\epsilon^2}{12}, 0)$ to ensure convexity. 
    A thorough investigation on recovering $\Kd$ from quantized data is still warranted. 
\end{rem}

\subsection{Quality vs. Quantity Trade-off} 

Another interesting avenue to pursue could be understanding the trade-off between the quality and quantity of the data. 
It has already been established that the estimation quality of the operator typically improves as the amount of (unquantized) data increases. 
In this work, we demonstrate that the estimation quality also improves with the word length of the quantization.
A higher word length per quantized data implies a smaller amount of data can be processed (i.e., communicated).
This, in turn, negatively affects the estimation quality while the higher quantization resolution is attempting to improve the estimation quality. 
Understanding this trade-off between quantity and quality of data would provide valuable insights for practical implementations under communication constraints.


\section{Conclusions} \label{sec:conclusions}
In this work, for the very first time, we investigate the effects of dither quantization on EDMD. 
In a large data regime, we show that the EDMD optimization problem with quantized data may be interpreted as a regularized EDMD optimization with unquantized data. 
We leveraged the law of large numbers to prove our claim. 
We also investigated the connection between the estimates of the Koopman operator under quantized and unquantized data in a finite data regime. 
We show that the quantized estimates converge to the true estimate as the quantization resolutions approach zero.

\section*{Acknowledgements}
The authors thank Indranil Nayak for generating the flow past cylinder data using MATLAB FEAtool.

\bibliographystyle{ieeetr}
\bibliography{references}

\appendix

\subsection{Some useful technical results}
In this section we provide some technical results that are used in the proof of Theorem~\ref{thm:equivalence}. 
Although similar results may be derived from textbook knowledge, we provide these proofs for completeness.

\begin{lm}\label{lem:cross_error}
    Let $e^i_t = \tilde\varphi^i(x_t) - \varphi^i(x_t)$ be the quantization error of the $i$-th observable at time step $t$. 
    Then, 
    \begin{align}
        \E[e^i_t e^j_s] = \begin{cases}
            \frac{\epsilon^2}{12}, \qquad i= j  ~~\mathrm{and}~~ t =s,\\
            0, \qquad~~ \mathrm{otherwise}.
        \end{cases}
    \end{align}
\end{lm}
\begin{proof}
    Due to the dither quantization scheme and the dither noise being uniform in $\left[ -\frac{\epsilon}{2}, \frac{\epsilon}{2} \right]$, each $e^i_t$ is independent and also uniformly distributed between $\left[ -\frac{\epsilon}{2}, \frac{\epsilon}{2} \right]$. 
    Consequently, when $i\ne j$ or $t\ne s$, we have $\E[e^i_t e^j_s] = \E[e^i_t] \E[e^j_s] = 0$. 
    On the other hand, $\E[(e^i_t)^2] = \frac{1}{\epsilon}\int_{-\epsilon/2}^{\epsilon/2} e^2 \mathrm{d}e = \frac{\epsilon^2}{12}.$
\end{proof}

\begin{corr} \label{corr:yij}
    For any fixed $i,j \in \{1,\ldots, N\}$, let $y^{ij}_t = e^i_t e^j_t$, where $i\ne j$. 
    Then,
    \begin{align}
         \lim_{T\to \infty}\frac{1}{T}\sum_{t=0}^{T-1} y^{ij}_t = 0.
    \end{align}
\end{corr}

\begin{proof}
    One may notice that $\{y^{ij}_t\}_{t\ge 0}$ is an i.i.d. sequence of random variables with $\E[y^{ij}_t] = 0$ (due to Lemma~\ref{lem:cross_error}) and $\E[(y^{ij}_t)^4] < \infty$. 
    Therefore, from strong law of large numbers, $\frac{1}{T}\sum_{t=0}^{T-1} y^{ij}_t \to 0$ almost surely. 
\end{proof}

\begin{corr} \label{corr:zij}
    For any fixed $i,j \in \{1,\ldots, N\}$, let $z^{ij}_t = e^i_t e^j_{t+1}$, where $i\ne j$. 
    Then,
    \begin{align}
         \lim_{T\to \infty}\frac{1}{T}\sum_{t=0}^{T-1} z^{ij}_t = 0.
    \end{align}
\end{corr}

\begin{proof}
    The proof follows the same steps as in the proof of \Cref{corr:yij}.
\end{proof}

\begin{corr} \label{corr:zi}
    Let $z^i_t = e^i_t e^i_{t+1}$. Then, 
    \begin{align}
         \lim_{T\to \infty}\frac{1}{T}\sum_{t=0}^{T-1} z^{i}_t = 0.
    \end{align}
\end{corr}
\begin{proof}
    Although $\{z^i_t\}_{t\ge 0}$ is an identically distributed sequence, it is not independent. 
    Therefore, the standard strong law of large numbers does not apply readily. 
    
    To proceed with the proof, let us first note that, for all $\tau \ge 1$,
    \begin{align} \label{eq:uncorrelated}
        \E[z^i_t z^i_{t+\tau}] &= \E[e^i_t e^i_{t+1}e^i_{t+\tau}e^i_{t+\tau+1}] \nonumber \\
         & = \E[e^i_t] \E[e^i_{t+1}e^i_{t+\tau}e^i_{t+\tau+1}] = 0,
    \end{align}
    where the second inequality follow from the fact that $e^i_t$ is independent of $e^i_{t+1}e^i_{t+\tau}e^i_{t+\tau+1}$ for all $\tau \ge 1$. 
    Therefore, \eqref{eq:uncorrelated} proves pairwise uncorrelation of the sequence $\{z^i_t\}_{t\ge 0}$. 

    Now let us define the random variable $\vartheta_T = \sum_{t=0}^{T-1} z^i_t$. 
    We note that, 
    \begin{align*}
        \E[(\vartheta_T)^4] &= \E\left[ \!\bigg( \sum_{t=0}^{T-1} z^i_t \bigg)^{\!\! 4} \right] \\
        & = T \E[(z^i_0)^{4}] + 3T(T-1) \E[(z^i_0 z^i_1)^2],
    \end{align*}
    where we have used $\E[z^i_k (z^i_\ell)^3] = 0$ since $z^i_k$ and $z^i_\ell$ are uncorrelated for all $k\ne \ell$ and $\E[z^i_k] = 0$ for all $k$.
    Given that $e^{i}_t$ is uniformly distributed in $\left[ -\frac{\epsilon}{2}, \frac{\epsilon}{2} \right]$, there exists a $K < \infty$ such that 
    \begin{align*}
        K = 4 \max\{ \E[(z^i_0)^4],  \E[(z^i_0 z^i_1)^2]\},
    \end{align*}
    which then implies 
    \begin{align} \label{eq: KBound}
        \E[(\vartheta_T)^4] \le K T^2,
    \end{align}
    for all $T \ge 1$. 
    Next, we use this bound to show that $\frac{1}{T}\sum_{t=0}^{T-1} z^{i}_t  \to 0$ almost surely. 
    To that end, let us start with
    \begin{align*}
        \E \left[ \sum_{T \ge 1} \left( \frac{\vartheta_T}{T} \right)^4\right] = \sum_{T \ge 1} \E \left[ \left( \frac{\vartheta_T}{T} \right)^4 \right] \le \sum_{T \ge 1} \frac{K} {T^{2}} < \infty, 
    \end{align*}
    where the first equality follows from the Fubini-Tonelli Theorem, and the first inequality follows from \eqref{eq: KBound}.
    Having proven that $\E \left[ \sum_{T \ge 1} \left( \frac{\vartheta_T}{T} \right)^4\right] < \infty$, we may conclude that $\sum_{T \ge 1} \left( \frac{\vartheta_T}{T} \right)^4 < \infty$ almost surely.
    Therefore, since the series converges, the underlying sequence must converge to zero, which implies 
    \begin{align*}
        \left( \frac{\vartheta_T}{T} \right)^4 \to 0 \qquad \text{almost surely}. 
    \end{align*}
    Consequently, we may conclude that 
    \begin{align*}
        \frac{1}{T}\sum_{t=0}^{T-1} z^{i}_t = \frac{\vartheta_T}{T} \to 0 \qquad \text{almost surely}. 
    \end{align*}
    This concludes the proof.
\end{proof}

\subsection{Proof of \Cref{thm:equivalence}} \label{AP:thm:equivalence} 
Notice that we may write 
\begin{align*}
     \| \tilde \Phi' - A \tilde \Phi\|^2 =  \sum_{t=0}^{T-1}\| \tilde \varphi(x_{t+1}) - A \tilde \varphi(x_t)\|^2.
\end{align*}
Now, let us define $e^i_t \triangleq \tilde\varphi^i(x_t) - \varphi^i(x_t)$ to be the quantization error on the $i$-th observable at time $t$, which is a zero-mean i.i.d. process due to the dither quantization. 
This implies $\E[e^i_t e^j_s] = 0$ when $i\ne j$ or $t\ne s$. 
Let us further define $e_t = [e^1_t,\ldots, e^N_t]^\top$. 

We may expand $\| \tilde \varphi(x_{t+1}) - A \tilde \varphi(x_t)\|^2$ as follows 
\begin{align} \label{eq:tilde_expansion}
    \| \tilde \varphi(x_{t+1}) & - A \tilde \varphi(x_t)\|^2 =  \|  \varphi(x_{t+1}) - A  \varphi(x_t) + e_{t+1} - Ae_t \|^2 \nonumber \\ \nonumber
     = &~ \|  \varphi(x_{t+1}) - A  \varphi(x_t)\|^2 + \| e_{t+1} \|^2 + \|A e_t\|^2 \\ \nonumber
     &~~ - 2e_{t+1}^\top A e_t + 2 e_{t+1}^\top (\varphi(x_{t+1}) - A  \varphi(x_t)) \\
     &~~ - 2 e_t^\top A^\top (\varphi(x_{t+1}) - A  \varphi(x_t)).
\end{align}
Now recall that $\{e^i_t\}_{t=  0:T}^{i = 1:N}$ is an i.i.d sequence, therefore, using the \textit{law of large numbers}, we may write 
\begin{align}
    \lim_{T\to \infty}\frac{1}{T}\sum_{t=0}^{T-1} e^i_t = 0
\end{align}
for all $i$, with almost sure probability. 
Similarly, we may also write 
\begin{align*}
    \lim_{T\to \infty} \frac{1}{T} \sum_{t=0}^{T-1}\|e_{t+1}\|^2 = \E[\|e_0\|^2] = \sum_{i=1}^N \E[(e^i_0)^2] \overset{(\dagger)}{=} N \frac{\epsilon^2}{12}
\end{align*}
almost surely, where $(\dagger)$ follows from \Cref{lem:cross_error}. 
Similarly, 
\begin{align*}
    \lim_{T\to \infty} \frac{1}{T} \sum_{t=0}^{T-1} e_{t+1}^\top A e_t = 0, 
\end{align*}
almost surely, due to \Cref{corr:zij} and \Cref{corr:zi}.

Therefore, for a large enough $T$, we may write
\begin{align} \label{eq:mainEquation}
    \frac{1}{T}\sum_{t=0}^{T-1} \| \tilde \varphi(x_{t+1})  - A \tilde \varphi(x_t)\|^2 = & \frac{1}{T} \sum_{t=0}^{T-1} \| \varphi(x_{t+1}) - A  \varphi(x_t)\|^2  \nonumber \\
    &+ N \frac{\epsilon^2}{12} + \frac{\epsilon^2}{12} \|A\|^2, 
\end{align}
where we have used $\frac{1}{T}\sum_{t = 0}^{T-1} e_t = \E[e] = 0$ on the last two terms of \eqref{eq:tilde_expansion}. 
Thus, 
\begin{align*}
    &\Kdt = \argmin_A  \| \tilde \Phi' - A \tilde \Phi\|^2 \\
    &= \argmin_A \frac{1}{T}\sum_{t = 0}^{T-1} \| \tilde \varphi(x_{t+1})  - A \tilde \varphi(x_t)\|^2 \\
    & =  \argmin_A \frac{1}{T} \sum_{t = 0}^{T-1} \| \varphi(x_{t+1}) - A  \varphi(x_t)\|^2 + \frac{\epsilon^2}{12} \|A\|^2. 
\end{align*}
This concludes the proof of \Cref{thm:equivalence}.

\subsection{Proof of \Cref{thm:K_epsilon}} \label{AP:thm:K_epsilon}
The closed form solution to the DMD problem in \eqref{eq:EDMD_quantized} with quantized data  is 
\begin{align} \label{eq:KDt_solution}
    \Kdt = \tilde\Phi' \tilde \Phi^\top \big( \tilde \Phi \tilde \Phi^\top\big)^{-1}. 
\end{align}
Let us denote $\tilde \Phi = \Phi + \Phi_\epsilon$, where the $ij$-th element of $\Phi_\epsilon$ is $e^i_j$, which is the quantization error corresponding to the $i$-th observable at time-step $j$. 
Given that $e^i_j$ is uniformly distributed in the range $\big[-\frac{\epsilon}{2}, \frac{\epsilon}{2} \big]$, one may conclude that $\|\Phi_\epsilon\| = O(\epsilon)$, where $O(\cdot)$ is the Big-O notation.
Similarly, we may also define $\tilde \Phi' = \Phi' + \Phi'_\epsilon$ and conclude that $\|\Phi'_\epsilon\| = O(\epsilon)$. 

After substituting $\tilde \Phi = \Phi + \Phi_\epsilon$ and $\tilde \Phi' = \Phi' + \Phi'_\epsilon$  in \eqref{eq:KDt_solution}, and after some simplifications, we obtain
\begin{align*}
    \Kdt & = \Kd - \Kd\big(  \Phi  \Phi^\top\Psi_\epsilon^{-1} + I  \big)^{-1} + \Gamma_\epsilon \big( \tilde \Phi \tilde \Phi^\top\big)^{-1} ,
\end{align*}
where $\Psi_\epsilon = \Phi_\epsilon \Phi^\top + \Phi \Phi_\epsilon^\top + \Phi_\epsilon \Phi_\epsilon^\top$  and $\Gamma_\epsilon = \Phi'_\epsilon \Phi^\top + \Phi' \Phi_\epsilon^\top + \Phi'_\epsilon \Phi_\epsilon^\top$. 
Therefore, we may write 
\begin{align*}
    \Kdt = \Kd + \K_\epsilon,
\end{align*}
where $\K_\epsilon =  \Gamma_\epsilon \big( \tilde \Phi \tilde \Phi^\top\big)^{-1} - \Kd\big(  \Phi  \Phi^\top\Psi_\epsilon^{-1} + I  \big)^{-1}$. 
The theorem is proven once we have shown that $\|\K_\epsilon\| = O(\epsilon)$. 
To that end, let us note that $\|\Psi_\epsilon\| = O(\epsilon)$ and $\|\Gamma_\epsilon\| = O(\epsilon)$, and therefore, $\|\K_\epsilon \| = O(\epsilon)$. 
This concludes the proof.

\subsection{Dataset generation for flow past cylinder}\label{AP:FPC}
\begin{figure} [h]
    \centering
      \includegraphics[width=1\linewidth]{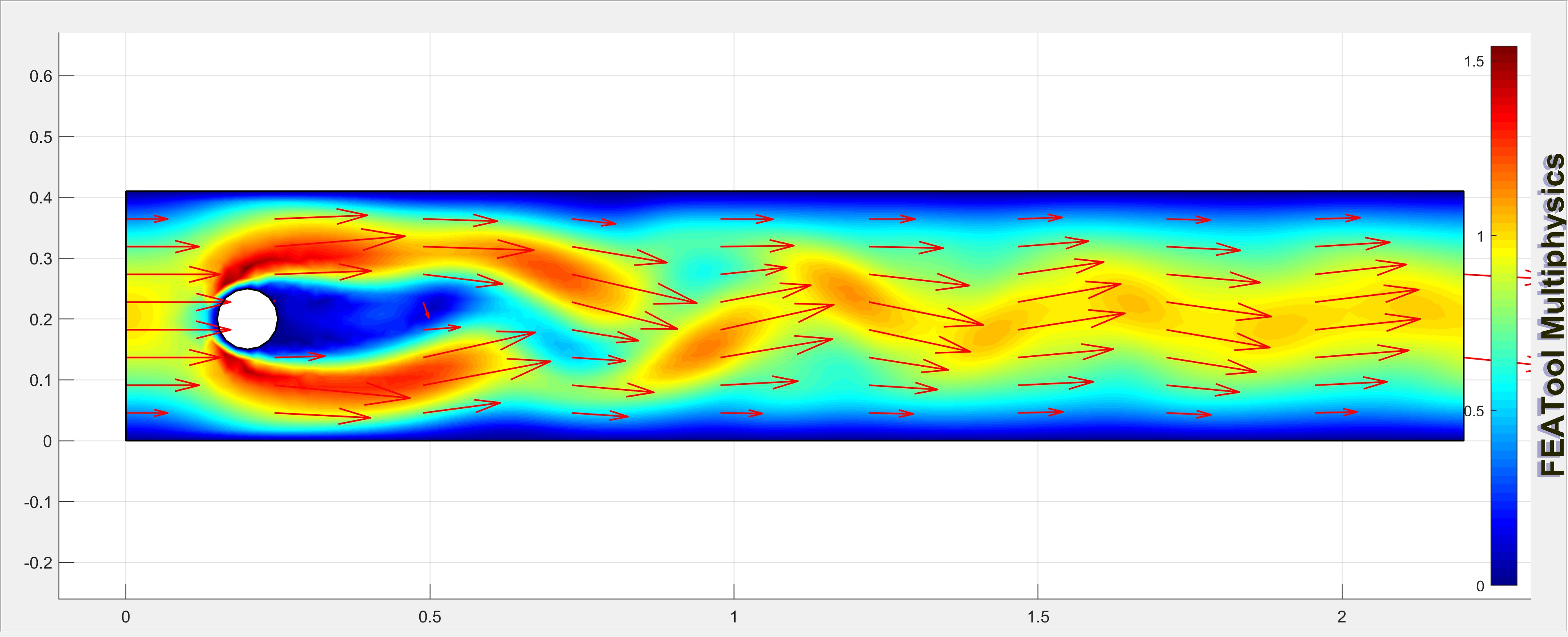}
  \caption{\small{Snapshot of velocity field at $t=80$s.}}\label{fig:flow_cyl_snap}
\end{figure}

The simulation setup \cite{Nayak2024} is shown in Fig.~\ref{fig:flow_cyl_snap}. 
The 2-D solution domain $(2.2~\text{m} \times 0.41~\text{m}) $ is discretized using irregular triangular mesh with number of nodes $N_0=2647$, and number of elements $N_2=5016$. The cylinder has the diameter of $0.1$ m with its center located at $(0.2~\text{m},0.2~\text{m})$. The flow is assumed to be incompressible, and governed by the Navier-Stokes equations with $\mathbf{u}$, $\mathbf{v}$ denoting the horizontal and vertical component of the velocity respectively while $\mathbf{p}$ denotes the pressure field. The density of the fluid is $\rho=1$ Kg/m$^3$ and its dynamic viscosity $\mu=0.001$ Kg/m s. 
The flow is unsteady with a maximum velocity of 1 m/s and mean velocity $\frac{2}{3}$ of the maximum velocity. The simulation starts with initial conditions $\mathbf{u}_0=\mathbf{v}_0=0$ and $\mathbf{p}_0=0$. The leftmost boundary is set as an inlet with a parabolic velocity profile, representing a fully developed laminar flow at the inlet. The rightmost boundary is set as an outflow (pressure boundary), where we specify the pressure but do not specify the velocity, allowing the flow to exit naturally based on the internal flow field. All other boundaries are treated as walls with a no-slip condition, i.e., the fluid velocity at the walls is zero. The simulation runs for a total of $80$ seconds, with a time-step size of $0.01$ seconds.

\end{document}